\documentclass[journal]{IEEEtran}

\usepackage{lipsum,amsmath}
\usepackage{bbm}
\usepackage{bm}
\usepackage{amsmath}
\usepackage{graphicx}
\usepackage{subfigure}
\usepackage{algpseudocode}
\usepackage{algorithmicx,algorithm}
\usepackage{algpseudocode}
\usepackage{amssymb}
\usepackage{float}
\usepackage{verbatim}
\usepackage{multirow}
\usepackage{makecell}
\usepackage{graphics}
\usepackage{diagbox}
\usepackage{xcolor}
\usepackage{multirow,booktabs}
\usepackage{setspace}
\usepackage{pifont}

\usepackage{cleveref}

\usepackage[numbers,sort&compress]{natbib}

\begin{document}

\title{Unsupervised Equivalent Contrastive Learning for Radio Signal Recognition}

\author{Shilian~Zheng,
        Jie~Chen,  
        Luxin~Zhang,
        and Xiaoniu~Yang

\thanks{(Shilian Zheng and Jie Chen are co-first authors.)(Corresponding author: Shilian Zheng.)}
\thanks{S. Zheng, J. Chen, and L. Zhang, X. Yang are with National Key Laboratory of Electromagnetic Space Security, Jiaxing 314033, China (e-mail: lianshizheng@126.com, chenjie91207@163.com, lxzhangMr@126.com, yxn2117@126.com).}

}

\maketitle

\begin{abstract}
Robust radio signal recognition is fundamental to spectrum management, electromagnetic space security, and intelligent wireless applications, yet existing deep-learning methods rely heavily on large labeled datasets and struggle to capture the multi-domain characteristics inherent in real-world signals. To address these limitations, we propose an unsupervised equivalent contrastive learning method that leverages four information-lossless equivalent transformations, spanning the time, instantaneous, frequency, and time–frequency domains, to construct multi-view and semantically consistent representations of each signal. An equivalent contrastive learning strategy then aligns these complementary views to learn discriminative and transferable embeddings without requiring labeled data. Once pre-training is completed, the resulting model can be directly fine-tuned on downstream tasks using only raw signal samples, without reapplying any equivalent transformations, which reduces computational overhead and simplifies deployment. Extensive experiments on four public datasets demonstrate that the proposed method consistently outperforms state-of-the-art contrastive baselines under linear evaluation, few-shot semi-supervised learning, and cross-domain transfer settings. Notably, the learned representations yield substantial gains in few-shot regimes and challenging channel conditions, confirming the effectiveness of multi-domain equivalent modeling in enhancing robustness and generalization. This work establishes a principled pathway for exploiting massive unlabeled radio data and provides a foundation for future self-supervised learning frameworks in wireless systems.
\end{abstract}


\section{Introduction}
\IEEEPARstart{T}{he} The electromagnetic spectrum serves as a critical yet invisible infrastructure of modern society, supporting nearly every aspect of human life \cite{9711564,8758230}. 
It supports large-scale information exchange in mobile, Wi-Fi, and satellite communication systems, as well as global navigation, meteorological observation, traffic control, and diverse radar and remote-sensing platforms for early warning and situational awareness \cite{chaccour2022seven}. Fundamentally, these applications can all be viewed as processes of transmitting, receiving, and exploiting radio signals within the electromagnetic space. As a result, the reliability of communication, the accuracy of navigation, and the effectiveness of monitoring and early-warning functions depend on our capability to acquire, process, and interpret radio signals \cite{11175176,10485272}. In this context, Robust radio signal recognition therefore plays an essential role in the signal processing pipeline, provides the fundamental support required for refined spectrum management, electromagnetic space security, and the advancement of intelligent electromagnetic applications \cite{liang2011cognitive,he2023channel}.

Traditional radio signal recognition methods, including feature-based \cite{dobre2007survey,4600222} and likelihood-based \cite{5351708,zhu2018likelihood} methods, rely on expert prior knowledge and manually engineered features. These methods are often limited by insufficient representational capacity, restricted generalization ability, and inadequate robustness in complex and dynamic electromagnetic environments. Recent advances in deep learning have led to a shift from knowledge-driven signal recognition toward data-driven approaches. By leveraging powerful automatic feature extraction and hierarchical pattern learning capabilities, deep-learning–based methods \cite{TU202235,o2018over,o2016convolutional,9106397,10146312} can directly learn discriminative representations from raw signals, thereby enabling end-to-end recognition without hand-crafted features and yielding substantial gains in accuracy, robustness, and adaptability. However, deep learning typically rely heavily on large-scale labeled samples, whereas obtaining high-quality annotated radio signals in real-world scenarios is often prohibitively expensive or even infeasible. In contrast, vast amounts of unlabeled radio signal data can be readily collected, making it a critical challenge to effectively exploit unlabeled data in the absence of annotations and thereby enhance modeling performance. Self-supervised learning \cite{10559458} provides a promising solution to this challenge. By designing proxy tasks during the pretraining stage, self-supervised learning leverages unlabeled data to learn robust feature representations, which can then be efficiently fine-tuned on downstream tasks. Consequently, it offers inherent advantages in data-scarce and cross-domain modeling scenarios.

Contrastive learning \cite{hu2024comprehensive} has emerged as as a dominant paradigm in self-supervised learning. It maximizes agreement between augmented views of the same signal while enforcing separation from others, thereby learning highly discriminative feature representations. Numerous studies have employed contrastive learning and its extended paradigms for pre-training, including SimCLR \cite{pmlr_v119_chen20j}, MoCo \cite{He_2020_CVPR}, DINO \cite{Caron_2021_ICCV}. In recent years, researchers have increasingly applied contrastive learning methods to radio signal recognition tasks \cite{10093837}. A typical approach designs diverse data augmentation strategies for the original signals, such as introducing noise perturbations \cite{10857965}, applying signal transformations \cite{10382665}, and performing cross-domain feature mapping \cite{li2025multi}, to generate multi-view positive and negative sample pairs. Then, deep learning models, including convolutional neural networks \cite{9451544} and transformers \cite{vaswani2017attention}, are used to extract multi-scale feature representations, which are optimized through contrastive learning in the feature space. Despite these advances, existing contrastive methods for radio signals face several limitations. First, frequency-domain information is often underutilized. Radio signals carry informative spectral cues such as harmonic patterns, spectral sparsity, and energy distribution. However, many existing methods mainly enforce time-domain consistency, which can lead to suboptimal extraction of spectral features and thus incomplete representations. Second, amplitude and phase variations introduce substantial uncertainty. Under varying transmission power, propagation distances, and channel conditions, signals of the same class can exhibit significant variations in amplitude and phase. Without explicitly modeling such invariances, the learned representations are prone to capturing irrelevant pseudo-features, which degrades generalization performance. Finally, multi-scale temporal structures are insufficiently captured. Radio signals often contain both short-term transient patterns and long-term statistical characteristics. Relying solely on a single time scale to construct contrast targets makes it difficult to obtain high-quality feature representations that are robust to different temporal resolutions.

To overcome these limitations, we propose an unsupervised equivalent contrastive learning method for radio signal recognition. The proposed method constructs multi-view equivalent representations by introducing a set of information-lossless equivalent transformations and performs equivalent contrastive learning strategy upon these representations, thereby learning transferable representations from unlabeled radio signals. Specifically, we design an equivalent representation generation module that applies four information-lossless equivalent transformations to radio signals to produce domain-specific multi-view representations in the time, instantaneous, frequency, and time–frequency domains, including time-domain augmentation, amplitude–phase decoupling, the fourier transform, and empirical mode decomposition. Subsequently, an equivalent contrastive learning strategy organizes the multi-view representations into positive and negative sample pairs and performs cross-representation contrastive optimization to learn unified feature embeddings. After the pre-training stage, the method can provide high-quality feature embedding for radio signal recognition tasks and significantly improve the performance by requiring only the original IQ signals without performing additional equivalent transformations in the fine-tuning stage.

In summary, the main contributions of this paper are as follows:
\begin{itemize}
    \item We propose an unsupervised equivalent contrastive learning method for radio signal recognition, which introduces four information-lossless equivalent transformations to learn a unified feature representations from large-scale unlabeled radio signals. After pre-training, the framework provides high-quality feature representations for various downstream tasks without requiring additional equivalent transformations, thereby enabling efficient adaptation and significant performance improvements.
    \item We design an equivalent representation generation module that constructs four information-lossless equivalent transformations for radio signal from the time, instantaneous, frequency, and time–frequency domains, via time-domain augmentation, amplitude–phase decoupling, fourier transform, and empirical mode decomposition. These transformations produce multi-view and complementary signal representations, providing more informative features for subsequent contrastive learning.
    \item We propose an equivalent contrastive learning strategy that forms positive and negative pairs across multiple views generated by the equivalent representation module. By enforcing cross-view consistency and contrastive separation, the strategy learns features that are more discriminative and generalizable for radio signal recognition.
    \item We evaluate the proposed method on four publicly available datasets with modulation recognition as a representative downstream task. The experimental results demonstrate that the proposed method can be seamlessly adapted to various downstream tasks and achieves significant performance improvements over the baseline approaches.
    
\end{itemize}

The remainder of this paper is organized as follows. Section \ref{related work} provides a comprehensive review and analysis of the related work. Section \ref{problem formulation} formally defines the problem formulation. Section \ref{Method} presents a detailed description of the proposed method. Section \ref{Experiments} reports the experimental results and provides an in-depth discussion. Finally, Section \ref{conclusion} concludes the paper.

\section{Related Work}
\label{related work}
\subsection{Contrastive Learning}
Contrastive learning has emerged as a powerful paradigm for self-supervised representation learning, whose core idea is to minimize the feature distance between positive pairs while maximizing that of negative pairs, thereby enabling the learning of more discriminative and generalizable feature representations. In recent years, contrastive learning has achieved remarkable progress, giving rise to a series of representative methods. For instance, Chen \textit{et al.} \cite{pmlr_v119_chen20j} propose the SimCLR framework, which leverages large-scale data augmentation to generate diverse views and constructs contrastive loss functions based on a large number of positive and negative sample pairs, thereby significantly improving the separability of the learned feature space. He \textit{et al.} \cite{He_2020_CVPR} introduced Momentum Contrast (MoCo) framework, which adopts a momentum-updated encoder and a dynamically maintained negative sample queue to enable efficient and stable contrastive learning. Caron \textit{et al.} \cite{NEURIPS2020_70feb62b} propose Swapping Assignments between multiple Views (SwAV), which performs online clustering and prediction swapping across multiple views, thereby avoiding explicit dependence on negative samples. Building on these advances, researchers explore approaches that remove explicit negative samples. Grill \textit{et al.} \cite{NEURIPS2020_f3ada80d} propose Bootstrap Your Own Latent (BYOL), which predicts target representations through self-supervised learning and achieves high-quality feature representations without using negative pairs. Similarly, Chen \textit{et al.} \cite{Chen_2021_CVPR} propose Simple Siamese Networks (SimSiam), which eliminates both negative samples and momentum encoders by employing a lightweight Siamese architecture and leverages feature prediction to facilitate stable unsupervised learning. Furthermore, Caron \textit{et al.} \cite{Caron_2021_ICCV} introduce Self-Distillation with No Labels (Dino), which adopts a teacher–student framework and utilizes a self-distillation strategy to learn robust and transferable visual representations. Following this, research shifts toward reducing feature redundancy and improving representational efficiency. Zbontar \textit{et al.} \cite{pmlr_v139_zbontar21a} propose Barlow Twins, which introduces feature decorrelation into contrastive learning by maximizing the similarity between embeddings of augmented views while minimizing inter-feature redundancy, thereby enhancing representation quality and discriminability. He \textit{et al.} \cite{He_2022_CVPR} propose Masked Autoencoders (MAE), which masks a large portion of image patches and reconstructs the missing content, formulating an efficient self-supervised pretext task. This approach extends the frontier of contrastive learning and broadens its applicability to more challenging representation modeling scenarios.

\subsection{Contrastive Learning for Radio Signals}
Benefiting from recent advances in computer vision and natural language processing, contrastive learning has attracted increasing attention in radio signal processing. For example, Kong \textit{et al.} \cite{10093837} propose a Transformer-based contrastive semi-supervised learning framework. It performs self-supervised contrastive pretraining on unlabeled samples with temporal regularization for data augmentation and further designs a convolutional Transformer network that integrates convolutional embeddings, attention bias, and attention pooling to enhance feature modeling capability. Bai \textit{et al.} \cite{10382665} propose Cooperative Contrast Learning for Modulation Signals (CoCL-Sig), a collaborative contrastive learning method that leverages both sequence and constellation diagram modalities. It applies modality-level alignment and instance-level feature constraints to improve representation stability and enhance downstream performance. Xiao \textit{et al.} \cite{10562208} propose Masked Contrastive Learning with Hard Negatives (MCLHN), which incorporates semantic-preserving augmentation and a temporal masking encoder under limited labeled data scenarios, effectively improving the generalization capability of automatic modulation classification and maintaining superior performance even with extremely few labeled samples. Li \textit{et al.} \cite{li2025multi} propose a multi-representation domain attentive contrastive learning framework that extracts high-quality signal representations from unlabeled data via cross-domain contrastive learning. It integrates both inter-domain and intra-domain contrastive mechanisms to enhance cross-domain modulation feature modeling and employs a domain-attention module to dynamically select representation domains for improved adaptability. Du \textit{et al.} \cite{10857965} propose a noise-aware contrastive learning approach that enhances model transferability under low-SNR conditions by removing the constraint of paired inputs. It enables the use of arbitrary combinations of clean and noisy signals from the same class and incorporates noise-level estimation to improve robustness against uncertain noise environments.

Despite the progress achieved by the above methods in automatic modulation recognition, those methods still rely on specific single-view representations or limited augmentation strategies, resulting in insufficient feature diversity and an inability to comprehensively capture the intrinsic characteristics of wireless signals across different domains and multiple scales. Therefore, we introduces an equivalent transformation strategy with lossless information to generate diverse representations without altering essential signal features, thereby enriching the feature space and enhancing representation capability.

\section{Problem Formulation}
\label{problem formulation}
In a wireless communication system, the transmitted signal undergoes modulation, power amplification, and frequency conversion before propagating through a complex electromagnetic environment to the receiver. The received baseband signal \(s_r(n)\) can be expressed as:
\begin{equation}
s_r(n) = [s_t(n) * h(n)]e^{j(2\pi \Delta f_n + \theta_0)} + w(n),
\label{eq1}
\end{equation}
where \(s_t(n)\) is the transmitted signal, \(h(n)\) is the impulse response of the wireless channel, \(\Delta f\) is the frequency offset, \(\theta_0\) is the initial phase offset, \(w(n)\) is additive Gaussian white noise with mean \(0\) and variance \(\sigma_w^2\), \(*\) is the convolution operation, \(n = 1,\dots,L\), \(L\) is the signal length.

Unlike supervised methods that rely on large amounts of labeled data, unsupervised learning aims to automatically learn discriminative and generalizable feature representations from large-scale unlabeled wireless signal datasets. Given an unlabeled dataset \(\mathcal{D} = \{ x_1, x_2, \dots, x_M \}\), \(x_i\) is the \(i\)-th received signal, \(M\) is the total number of samples, the objective of unsupervised pre-training is to learn a feature extractor \(f_\theta(\cdot)\) parameterized by \(\theta\), which maps the original signals \(x_i\) into a low-dimensional latent feature:
\begin{equation}
z_i = f_\theta(x_i), \quad z_i \in \mathbb{R}^d.
\label{eq2}
\end{equation}
The features \(z_i\) are expected to possess strong discriminability, robustness, and generalization capability, thereby enabling the pre-trained model \(f_\theta(\cdot)\) to be efficiently fine-tuned on downstream datasets \(\mathcal{D}_{ft}\) without requiring retraining of the entire model.

\begin{figure*}[htbp]
\centering
\includegraphics[width=0.80\textwidth]{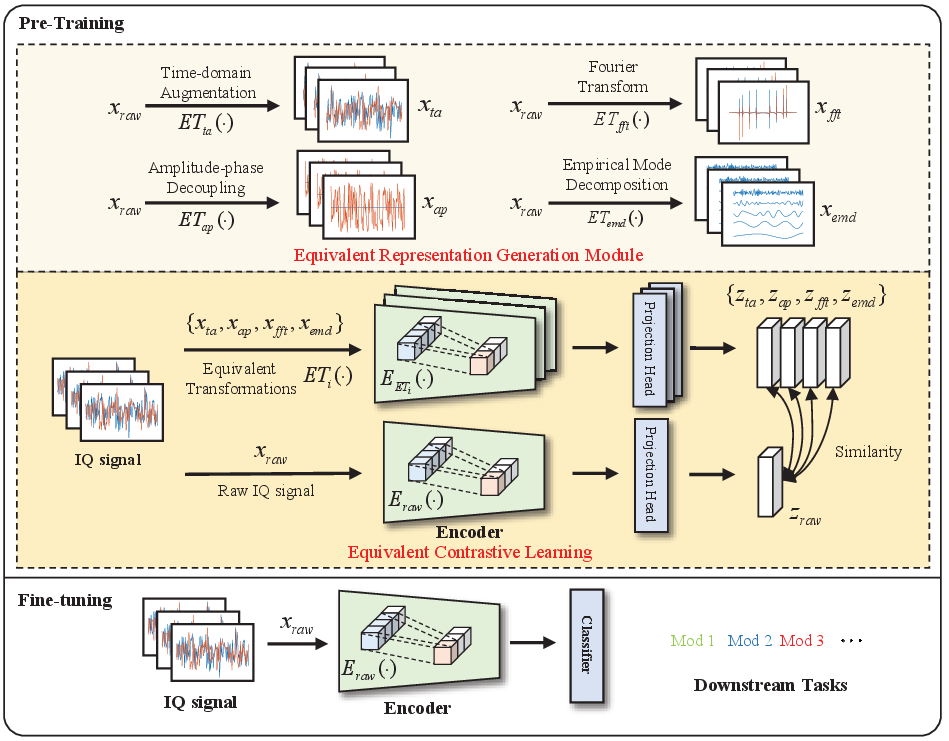}
\caption{The overall architecture of our proposed method.}
\label{fig1}
\end{figure*}

\section{Method}
\label{Method}
\subsection{Overall Framework}

The overall architecture of the proposed method is shown in Fig. \ref{fig1}. It comprises a pre-training stage and a fine-tuning stage. During pre-training, an equivalent-representation generation module generates four information-lossless equivalent representation of each radio signal capturing complementary feature representations in the time, instantaneous, frequency, and time–frequency domains. These multi-view representation are then used in an equivalent contrastive learning strategy that constructs positive pairs from different equivalent views of the same signal and negative pairs from different signals. By encouraging agreement among positive pairs while separating negative ones within a shared embedding space, the model learns feature representations with improved discriminability and robustness. After pre-training, the resulting encoder can be directly adapted to downstream radio recognition tasks using only the original IQ signals and without requiring any additional equivalent transformations, enabling efficient fine-tuning and substantial performance gains.

\subsection{Equivalent Representation}

\textbf{Definition 1.} Let \(x\) denote a radio signal. A mapping \(T: x \rightarrow x'\) is defined as an equivalent transformation if and only if there exists a unique inverse mapping \(T^{-1}: x' \rightarrow x\) such that \(T^{-1}(T(x)) = x\). Such transformations modify only the observable representation of the signal while strictly preserving its task-relevant properties.

In the proposed method, the equivalent representation generation module constructs multi-view representations through a set of equivalent transformations, enhancing cross-domain feature expressiveness while preserving the informational completeness of the raw signal. This design effectively addresses the limitations of existing radio signal representation methods, such as time-domain waveforms \cite{10857965}, which rely on a single-view representation and suffer from limited feature expressiveness. Based on Definition 1 of equivalent transformations, we design four types of equivalent transformations that extract key characteristics of radio signals from the time, instantaneous, frequency, and time–frequency domain. Each transformation admits a corresponding inverse mapping, enabling the original signal \(x_{{raw}}\) to be uniquely reconstructed from its transformed representation. This guarantees that no information is lost while obtaining diverse views, thereby establishing a solid foundation for the subsequent equivalent contrastive learning strategy.

\subsubsection{{Time Domain}}
The equivalent transformation in the time domain is rotation and time-cycle shift operation. First, we extract the real and imaginary parts of the original signal, which can be expressed as:
\begin{equation}
\begin{bmatrix}
\mathbf{I} \\
\mathbf{Q}
\end{bmatrix}
=
\begin{bmatrix}
\text{Real}(x_{{raw}}) \\
\text{Imag}(x_{{raw}})
\end{bmatrix}
=
\begin{bmatrix}
I_0,I_1,\dots,I_{L-1} \\
Q_0,Q_1,\dots,Q_{L-1} 
\end{bmatrix}.
\label{eq3}
\end{equation}
Then, the rotation operation can be expressed as:
\begin{equation}
x^{1}_{ta} = 
\begin{bmatrix}
\cos\alpha & -\sin\alpha \\
\sin\alpha & \cos\alpha
\end{bmatrix}
\begin{bmatrix}
\mathbf{I} \\
\mathbf{Q}
\end{bmatrix},
\quad \alpha \in (0, \pi],
\label{eq4}
\end{equation}
where \(\alpha\) is the rotation angle. The time cyclic shifting operation can be expressed as: 
\begin{equation}
x^{2}_{ta} = 
\begin{bmatrix}
I_i,\dots,I_{L-1},I_0,\dots,I_{i-2} \\
Q_i,\dots,Q_{L-1},Q_0,\dots,Q_{i-2} 
\end{bmatrix}, i\in(0,L-1],
\label{eq5}
\end{equation}
where \(i\) is the length of the shift. When performing time-domain augmentation, the time-domain transformations is selected with equal probability, and the final generated time domain representation can be expressed as:
\begin{equation}
x_{ta} = ET_{{ta}}(x_{{raw}}) \leftarrow \{{x^{1}_{ta},x^{2}_{ta}}\}.
\label{eq6}
\end{equation}

\subsubsection{{Instantaneous Domain}}
The equivalent transformation in the instantaneous domain is amplitude–phase decoupling. The amplitude contains information related
to energy and strength of the signal, while the phase reflects
its frequency content and temporal evolution. The original signal \(x_{{raw}}(n)\) can be represented as the sum of its real and imaginary parts:
\begin{equation}
x_{{raw}}(n) = I(n) + jQ(n),
\label{eq7}
\end{equation}
where \(I(n)\) and \(Q(n)\) is the in-phase and quadrature components, respectively. The instantaneous amplitude and phase of the signal can then be derived as:
\begin{equation}
\begin{aligned}
x_{amp}(n) &= \sqrt{ I^2(n) + Q^2(n) }, \\
x_{pha}(n) &= \arctan \left( \frac{Q(n)}{I(n)} \right).
\end{aligned}
\label{eq8}
\end{equation}
It is worth noting that this decoupling process is fully reversible. The original signal can be reconstructed from its amplitude and phase components as:
\begin{equation}
x_{raw}(n)=x_{amp}(n) e^{{j x_{pha}(n)}}.
\label{eq9}
\end{equation}
Finally, the generated instantaneous-domain representation can be represented as:
\begin{equation}
x_{ap} = ET_{ap}(x_{{raw}}) = 
\begin{bmatrix}
x_{amp}(n) \\
x_{pha}(n) 
\end{bmatrix}.
\label{eq10}
\end{equation}

\subsubsection{{Frequency Domain}}
The equivalent transformation in the frequency domain is Fourier Transform \cite{heckbert1995fourier}. The Fourier Transform is instrumental in characterizing the fundamental properties of a signal, such as its periodicity and frequency content. Specifically, the \(N\)-point discrete Fourier transform is defined by the following expression:
\begin{equation}
X(k)=\sum_{n=0}^{N-1} x(n)e^{-j \frac{2 \pi}{N}kn},k=0,\dots,N-1,
\label{eq11}
\end{equation}
where \(x(n)\) is the time-domain signal, and \(X(k)\) represents its corresponding frequency-domain counterpart. The Fourier Transform is instrumental in characterizing the fundamental properties of a signal, such as its periodicity and frequency content. Its inverse transform is:
\begin{equation}
x(n)=\frac{1}{N} \sum_{k=0}^{N-1} X(k)e^{j \frac{2 \pi}{N}kn}.
\label{eq12}
\end{equation}
As the discrete Fourier transform and its inverse form a bijective mapping, the transformation does not result in any information loss. Instead, it offers an alternative perspective of the signal by converting it from the time domain to the frequency domain. After applying the Fourier transform, the generated frequency domain representation can be expressed as:
\begin{equation}
x_{fft}=ET_{fft}(x_{{{raw}}})=
\begin{bmatrix}
\text{Real}(X(k)) \\
\text{Imag}(X(k))
\end{bmatrix}.
\label{eq13}
\end{equation}

\subsubsection{{Time–frequency Domain}}
The equivalent transformation in the time–frequency domain is empirical mode decomposition (EMD) \cite{huang1998empirical}. EMD effectively extracts the instantaneous frequency components of the signal and decomposes the complex signal into multiple frequency bands. This enables the model to focus on different spectral regions and capture multi-scale discriminative features, thereby enhancing the understanding of the intrinsic characteristics of the signal. Specifically, the original signal \(x_{{raw}}(n)\) can be represented as the combination of multiple intrinsic mode functions (IMFs) and a residual component, formulated as:
\begin{equation}
x_{raw}(n)=\sum_{i=1}^M \mathrm{IMF}_i(n)+\mathrm{Res}(n),
\label{eq14}
\end{equation}
where \(M\) is the number of IMFs, \(\mathrm{IMF}_i(n)\) represents the \(i\)-th intrinsic mode function, and \(\mathrm{Res}(n)\) corresponds to the residual component of the signal. By applying the empirical mode decomposition, the generated time–frequency representation can be expressed
as:
\begin{equation}
x_{emd}=ET_{emd}(x_{{{raw}}})=
\begin{bmatrix}
\mathrm{IMF}_1(n) \\
\dots \\
\mathrm{IMF}_M(n) \\
\mathrm{Res}(n)
\end{bmatrix}.
\label{eq15}
\end{equation}

After these four equivalent transformations, we obtain four equivalent representations \(\{x_{ta}, x_{ap}, x_{fft}, x_{emd}\}\) of the original signal \(x_{{raw}}\).

\subsection{Equivalant Contrastive Learning}
In the proposed method, the equivalant contrastive learning strategy leverages the multi-view equivalent representations generated by the equivalent transformations to learn a unified feature representation with enhanced discriminability and robustness. As shown in Fig. \ref{fig1}, each raw signal \(x_{raw}\) is first processed by the encoder \(E_{raw}(\cdot)\) to extract its representation in the raw feature space, and the projection head \(g_{raw}(\cdot)\) subsequently maps it into the contrastive embedding space to obtain the feature vector \(z_{raw}=g_{raw}(E_{raw}(x_{raw}))\). Meanwhile, each transformed view \(x_k\), generated from the same raw signal through different equivalent transformations, is processed by their corresponding transformation-specific encoders \(E_{ET_{k}}(\cdot)\) and projection heads \(g_k(\cdot)\) to produce the feature vectors \(z_{k}=g_{k}(E_{ET_k}(x_{k}))\), where \(k \in \{ta,ap,fft,emd\}\) correspond to the four equivalent representations in the time, instantaneous, frequency, and time-frequency domain, respectively.

During contrastive learning, the representations of the same signal from the raw view and an equivalent-transformed view are regarded as positive pairs \( (x_{raw}, x_k) \), while features from other samples are treated as negative pairs. Similarity between two embeddings is measured using the normalized cosine similarity:
\begin{equation}
\mathrm{sim}(z_i,z_j)=\frac{z_i^\top z_j}{\|z_i\| \|z_j\|}.
\label{eq16}
\end{equation}
Then, we aim to maximize the similarity between the same-sample representations across views while minimizing the distance between representations from different samples. 
For each transformed view \(k\), we align the raw-view embedding \(z_{raw}^i\) with its corresponding transformed embedding \(z_k^i\) using as follows:
\begin{equation}
\mathcal{L}_{raw,k} = -\frac{1}{B} \sum_{i=1}^B \mathrm{log} \left( \frac{\exp(\mathrm{sim}(z_{raw}^i,z^i_{k}) / \tau)}{ \sum_{j=1}^B \exp(\mathrm{sim}(z_{raw}^i,z^j_{k})/\tau)} \right),
\label{eq17}
\end{equation}
where \(B\) is the batch size and \(\tau\) is the temperature parameter. Because this formulation aligns only raw-to-transformed directions, it may lead to asymmetric embedding collapse. To ensure consistent alignment across transformations, we introduce a bidirectional objective that jointly maximizes the agreement between \((z_{raw}, z_k)\) and \((z_k, z_{raw})\). Integrating all four transformation branches, the overall equivalant contrastive loss is given by:
\begin{equation}
\mathcal{L} = \sum_{k}\frac{1}{2}(\mathcal{L}_{raw,k} + \mathcal{L}_{k,raw}).
\label{eq18}
\end{equation}
This bidirectional and multi-view loss encourages the model to develop a coherent embedding geometry that captures invariant structure across signal domains, ultimately yielding more robust and transferable representations.

\subsection{Fine-tuning}
After pre-training, the model acquires a strong ability to extract representations that are both discriminative and robust generalization. Notably, the proposed method does not require applying the equivalent transformations during the fine-tuning stage, and can be fine-tuned efficiently using raw IQ signals alone.

For a downstream radio signal recognition task, we consider a labeled dataset \(\mathcal{D}_{ft}=\{x_i,y_i\}_{i=1}^{M_{ft}}\), where \(M_{ft}\) denotes the number of samples and each label \(y_i \in \{1, \dots, C\}\) corresponds to one of the \(C\) signal classes. The goal of this task is to learn a mapping from raw IQ signal \(x_i\) to their corresponding signal class \(y_i\). During the fine-tuning stage, the raw signal encoder \(E_{raw}(\cdot)\) is reused to obtain latent representations \(z_i = E_{raw}(x_i)\), upon which a task-specific classifier \(h_{fc}(\cdot)\) (e.g., a single-layer or multi-layer fully connected network) is applied to produce the predicted class \(\hat{y}=h_{fc}(z_i)\). The model parameters are optimized in a supervised manner using the standard cross-entropy loss

\begin{equation}
\mathcal{L}_{ft} = -\frac{1}{B}\sum_{i=1}^{B} \sum_{c=1}^{C}
y_{i,c} \log (\hat{y}_{i,c}),
\label{eq19}
\end{equation}
where \(B\) is the batch size, \(y_{i,c}\) is the ground-truth probability belonging to the \(c\)-th class, \(\hat{y}_{i,c}\) is the predicted probability belonging to the \(c\)-th class. Depending on the requirements of the downstream task, the training process may either update only the classifier parameters \(h_{fc}(\cdot)\), or jointly optimize both the encoder \(E_{raw}(\cdot)\) and the classifier \(h_{fc}(\cdot)\) through end-to-end fine-tuning to achieve stronger task-specific adaptation.

\section{Experiments}
\label{Experiments}
\subsection{Experimental Setup}
\subsubsection{Dataset} We evaluate the proposed method on several open-source benchmark datasets.
\begin{itemize}
    \item RML2016.10A \cite{o2016convolutional}: The dataset contains 11 modulation types, including digital modulations such as BPSK, QPSK, 8PSK, 16QAM, 64QAM, BFSK, CPFSK, and PAM4, as well as analog modulations such as WBFM, AM-SSB, and AM-DSB. The signal-to-noise ratio (SNR) of each type ranges from -20 dB to 18 dB, with a step size of 2 dB. The length of each signal is 128. Each type at each SNR contains 1,000 samples, totaling 220,000 samples.
    \item HKDD\_AMC12 \cite{10042021}: The dataset contains 12 modulation types, including BPSK, QPSK, 8PSK, OQPSK, 16QAM, 32QAM, 64QAM, 4PAM, 8PAM, 2FSK, 4FSK, and 8FSK. The SNR of each type ranges from -20 dB to 18 dB, with a step size of 2 dB. The length of each signal is 512. The training set contains 1,000 samples for each type at each SNR, and the test set includes 500 samples, for a total of 252,000 and 126,000 samples respectively.
    \item HKDD\_AMC36 \cite{10042021}: The dataset contains 36 modulation types, including BPSK, QPSK, 8PSK, OQPSK, 16PSK, 32PSK, 2FSK, 4FSK, 8FSK, 16QAM, 32QAM, 64QAM, 128QAM, 256QAM, 16APSK, 32APSK, 64APSK, 128APSK, 256APSK, 4PAM, 8PAM, 16PAM, MSK, GMSK, 4CPM, 8CPM, OFDM-BPSK, OFDM-QPSK, OFDM-16QAM, AM, FM, OOK, 4ASK, 8ASK, AMMSK, and FM-MSK. The SNR of each type ranges from -20 dB to 30 dB, with a step size of 2 dB. The length of each signal is 1024. The training set contains 1,000 samples for each type at each SNR, and the test set includes 500 samples, for a total of 936,000 and 468,000 samples respectively.
    \item HisarMod2019.1 \cite{zhang2022deep}: The dataset contains 26 modulation types, including digital modulations such as 2FSK, 4FS, 8FSK, 16FSK, 4PAM, 8PAM, 16PAM, BPSK, QPSK, 8PSK, 16PSK, 32PSK, 32PSK, 4QAM, 8QAM, 16QAM, 32QAM, 64QAM, 128QAM, and 256QAM, as well as analog modulations such as AM-DSB, AM-SC, AM-USB, AM-LSB, FM, and PM. The SNR of each type ranges from -20 dB to 30 dB, with a step size of 2 dB. The length of each signal is 1024. The training set contains 1,000 samples for each type at each SNR, and the test set includes 500 samples, for a total of 520,000 and 260,000 samples respectively.
\end{itemize}

\subsubsection{Implementation Detials}
In the pre-training phase, we employ the Adam optimizer with an initial learning rate of 0.001, decreasing by 0.2 every 10 epochs for a total of 80 epochs. In the fine-tuning phase, we also use the Adam optimizer, but with an initial learning rate of 0.0005 and a training period of 60 epochs. The batch size is set to 128 for both pretraining and fine-tuning. All algorithms are implemented in Python using the PyTorch framework.

In the fine-tuning stage, we adopt three evaluation strategies to verify the effectiveness of self-supervised pretraining: linear evaluation (LE), semi-supervised learning (SSL), and transfer learning (TL). In LE, we freeze the pretrained feature extractor and train only a linear classifier on top of the extracted representations to assess feature quality. In SSL, we fine-tune the pretrained model with a small number of labeled samples while keeping the pre-trained weights initialized, evaluating the performance in few-shot scenarios. In TL, we transfer the pretrained model to a new dataset and apply both LE and SSL to comprehensively evaluate the model’s cross-dataset generalization and transfer capabilities.

\begin{table*}[h]
\small
\renewcommand\arraystretch{1.3}
\centering
\caption{The result of linear evaluation.}
\begin{tabular}{lcccc}
\hline\hline
Method & RML2016.10A (\(\%\))  & HKDD\_AMC12 (\(\%\))  & HKDD\_AMC36 (\(\%\))  & HisarMod2019.1 (\(\%\)) \\ \hline
Supervised & 61.78 & 58.87 & 63.46 & 74.73 \\ 
SimCLR \cite{pmlr_v119_chen20j}  & \underline{47.74} & 34.98 & 40.99 & 31.85   \\
SimSiam \cite{Chen_2021_CVPR}  & 32.62 & 9.72 & 16.91 & 14.53 \\
Dino \cite{Caron_2021_ICCV} & 46.59 & \underline{36.72} & \underline{41.88} & \underline{34.94} \\ 
Our & \textbf{54.42} & \textbf{47.58} & \textbf{48.34} & \textbf{48.23} \\
\hline \hline 
\multicolumn{5}{l}{\textbf{bold} denotes the best, \underline{underline} denotes the second best.}
\end{tabular}
\label{tab1}
\end{table*}

\subsection{Comparison with State-of-the-Art}
We compare the proposed method with classic baseline methods, including SimCLR \cite{pmlr_v119_chen20j}, SimSiam \cite{Chen_2021_CVPR}, and DINO \cite{Caron_2021_ICCV}. All methods use the same backbone network, ResNexXt50 \cite{xie2017aggregated}, as the feature extractor to ensure a fair performance comparison between methods with similar feature extraction capabilities.

\subsubsection{Linear Evaluation}

\begin{figure}
    \centering
    \includegraphics[width=0.4\textwidth]{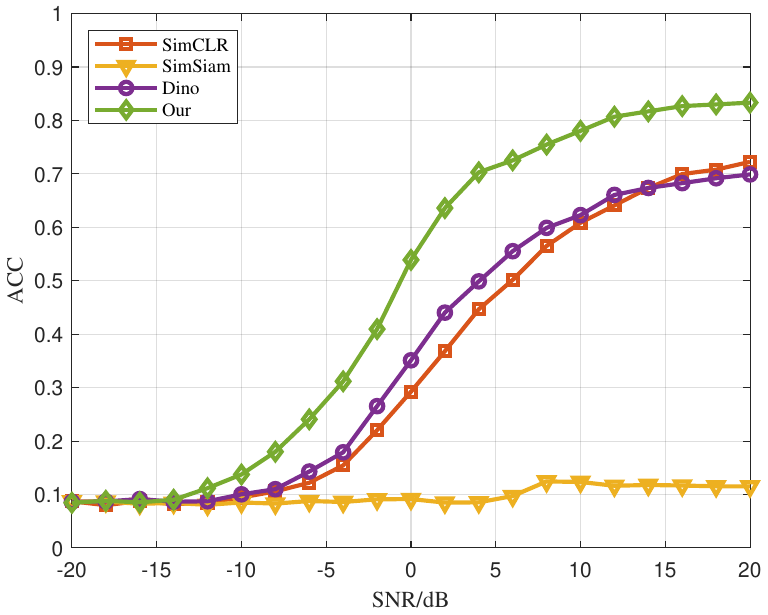}
    \caption{The performance of linear evaluation at different SNR level for dataset HKDD\_AMC12. }
    \label{fig2}
\end{figure}

Following standard practice, we adopt the widely used linear evaluation protocol \cite{He_2020_CVPR,li2025multi}, where the pretrained feature extractor is frozen and only a lightweigh classifier is trained on top of the extracted representations. Since the backbone is not updated during this process, linear evaluation provides an objective assessment of the intrinsic separability of the learned feature space. Table \ref{tab1} reports the performance of linear evaluation on different self-supervised methods and the supervised upper bound across four datasets.

As shown in Table \ref{tab1}, the proposed method consistently outperforms existing baselines on datasets RML2016.10A and HKDD\_AMC12 with small numbers of classes. On RML2016.10A, the supervised model achieves 61.78\%, while SimCLR, DINO, and SimSiam obtain 47.74\%, 46.59\%, and 32.62\%, respectively. In contrast, the proposed method reaches 54.42\%, surpassing SimCLR and DINO by 6.7\% and 7.8\%, and narrowing the gap to the supervised upper bound to approximately 7\%. A similar performance pattern is observed on HKDD\_AMC12. Moreover, Fig. \ref{fig2} further illustrates the performance differences across SNR level. When SNR \(\geq -10\) dB, the performance of the proposed method increases rapidly and remains consistently higher than SimCLR and DINO by 5\% \(\sim\) 20\% throughout the medium-to-high SNR range. In contrast, SimSiam almost completely fails on this dataset. These results collectively demonstrate that, under limited class diversity, the proposed method can reliably learn higher-quality representations than existing baselines.

On the more challenging HKDD\_AMC36 dataset, where the number of modulation classes is substantially larger and high-order modulations introduce severe feature entanglement, the proposed method continues to outperform the baselines, exceeding SimCLR and DINO by 7.4\% and 6.5\%, respectively. This indicates that the proposed framework retains strong inter-class separability even when modulation types become considerably more diverse. HisarMod2019.1 represents the most realistic setting among the four datasets, containing signals transmitted over five different wireless communications channels. In this complex scenario, SimCLR, SimSiam, and DINO achieve only 31.85\%, 14.53\%, and 34.94\%, respectively. In contrast, the proposed method achieves 48.23\%, outperforming SimCLR and DINO by 16.4\% and 13.3\%. This indicates that the representation learned by our method is more robust.

\subsubsection{Semi-supervised Learning}

\begin{table*}[h]
\small
\renewcommand\arraystretch{1.3}
\centering
\caption{The result of semi-supervised learning. }
\begin{tabular}{lcccccccc}
\hline\hline
Method & \multicolumn{2}{c}{RML2016.10A (\(\%\)) } & \multicolumn{2}{c}{HKDD\_AMC12 (\(\%\))} & \multicolumn{2}{c}{HKDD\_AMC36 (\(\%\))} & \multicolumn{2}{c}{HisarMod2019.1 (\(\%\))}  \\ 

\emph{Label faction:} & \(0.5\%\) & \(1\%\) & \(0.5\%\) & \(1\%\) & \(0.5\%\) & \(1\%\) & \(0.5\%\) & \(1\%\) \\ \hline
Supervised & 9.11 & 9.09 & 28.79 & 32.06 & 38.72 & 44.63 & \underline{27.93} & 28.29 \\ 
SimCLR \cite{pmlr_v119_chen20j}  & 40.45 & 42.09 & 31.41 & \underline{36.27} & \underline{46.87} & \underline{49.35} & 26.98 & \underline{28.43}   \\
SimSiam \cite{Chen_2021_CVPR}  & 41.77 & 46.61 & \underline{32.84} & 36.01 & 46.74 & 48.64 & 21.62 & 22.34  \\
Dino \cite{Caron_2021_ICCV} & \underline{43.21} & \underline{46.98} & 31.63 & 34.48 & 42.63 & 46.67 & 25.26 & 26.91\\ 
Our & \textbf{47.61} & \textbf{50.28} & \textbf{42.05} & \textbf{48.16} & \textbf{49.04} & \textbf{51.18} & \textbf{30.48} & \textbf{31.12} \\
\hline \hline 
\multicolumn{9}{l}{\textbf{bold} denotes the best, \underline{underline} denotes the second best.}
\end{tabular}
\label{tab2}
\end{table*}

\begin{figure*}
    \centering
    \subfigure[]{
    \label{fig3:a}
    \includegraphics[width=0.4\textwidth]{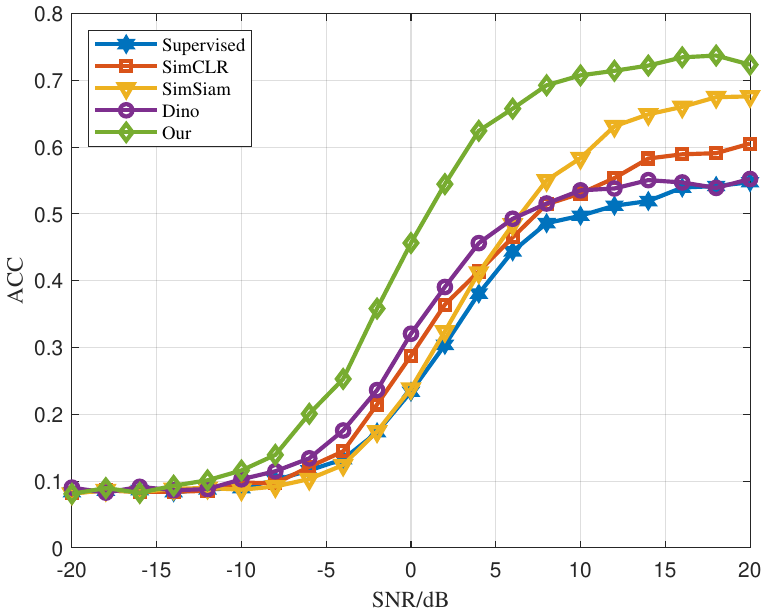}}
    \subfigure[]{
    \label{fig3:b}
    \includegraphics[width=0.4\textwidth]{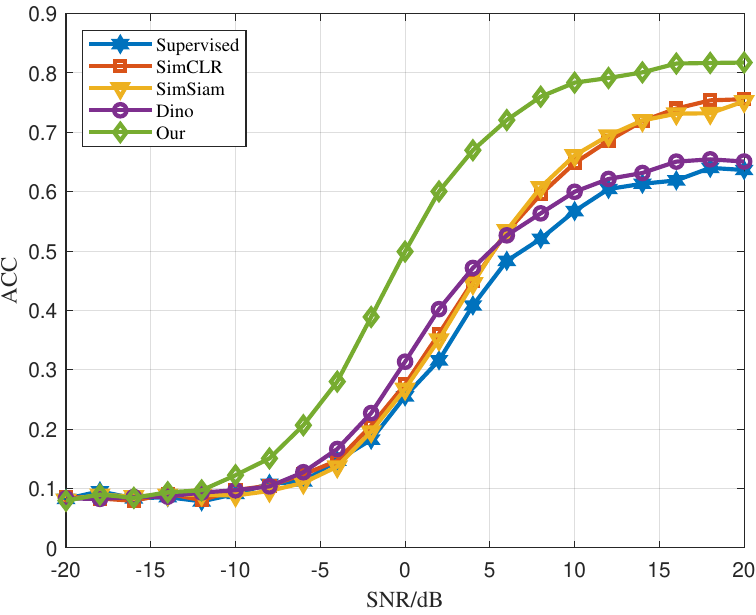}}
    \caption{The performance of semi-supervised learning in few-shot scenarios for dataset HKDD\_AMC12. (a) $0.5\%$ of samples, (b) $1\%$ of samples. }
    \label{fig3}
\end{figure*}

We follow \cite{zhai2019s4l} and sample 0.5\% and 1\% of labeled training datasets in a class-balanced way. The entire backbone is fine-tuned directly on this labeled subset without any additional regularization. The overall performance of all methods under different label fractions is reported in Table \ref{tab2}.

As shown in Table \ref{tab2}, the performance of the supervised model drops sharply when label availability is extremely restricted. With only 1\% of the labels, the supervised method on RML2016.10A, HKDD\_AMC12, and HKDD\_AMC36 are merely 9.09\%, 32.06\%, and 44.63\%, respectively. In contrast, the baseline methods (SimCLR, SimSiam, and DINO) show noticeable improvements over the supervised method, but the proposed method consistently achieves the best performance. Under the 1\% label condition, the proposed approach attains 50.28\%, 48.16\%, and 51.18\% on RML2016.10A, HKDD\_AMC12, and HKDD\_AMC36, outperforming the second-best method by approximately 3.3\%, 11.8\%, and 1.8\%, respectively, demonstrating substantially stronger sample efficiency. When the number of labeled samples is further reduced to 0.5\%, our method remains the best across all datasets, surpassing the second-best method by about 4.4\%, 9.2\%, and 2.2\%, respectively. Notably, on the HisarMod2019.1 dataset with more complex channel conditions, the performance of baseline methods even shows varying degrees of decline. For instance, with 0.5\% of the labels, SimCLR, SimSiam, and DINO drop by 0.9\%, 6.3\%, and 2.7\%, respectively. With 1\% of the labels, SimSiam and DINO further decline by 5.9\% and 1.3\%. In sharp contrast, the proposed method still improves over the supervised model by 2.5\% (0.5\% of labels) and 2.8\% (1\% of labels), indicating that our learned representations remain significantly more robust under highly non-ideal channel conditions.

In addition, Fig. \ref{fig3} illustrates the semi-supervised performance across different SNR levels on HKDD\_AMC12. When SNR \(\geq -6\) dB, the proposed method clearly outperforms all baseline methods under both labeling ratios, achieving advantages of approximately 5\% to 20\%. The superiority is even more significant in the low-to-medium SNR regime. For example, at SNR \(=4\) dB, our method surpasses the best baseline by 16.8\% (0.5\% of labels) and 19.83\% (1\% of labels).

Overall, these results demonstrate that the proposed method not only exhibits superior data efficiency but also offers substantially stronger noise robustness. Even under severely limited supervision and challenging channel conditions, it is able to learn highly discriminative representations, thereby providing a more reliable initialization for downstream tasks.

\subsubsection{Transfer Learning}

\begin{table*}[h]
\small
\renewcommand\arraystretch{1.3}
\setlength{\tabcolsep}{4.5pt}
\centering
\caption{The result of transfer learning with single source transfer setting. }
\begin{tabular}{lcccccccccccc}
\hline\hline
Source Dataset & \multicolumn{6}{c}{HKDD\_AMC36 (\(\%\))}  & \multicolumn{6}{c}{HisarMod2019.1 (\(\%\))}\\ \cmidrule(lr){2-7} \cmidrule(lr){8-13}
Target Dataset & \multicolumn{2}{c}{RML2016.10A  } & \multicolumn{2}{c}{HKDD\_AMC12} & \multicolumn{2}{c}{HisarMod2019.1 } & \multicolumn{2}{c}{RML2016.10A  } & \multicolumn{2}{c}{HKDD\_AMC12} & \multicolumn{2}{c}{HKDD\_AMC36 } \\ \cmidrule(lr){2-3} \cmidrule(lr){4-5} \cmidrule(lr){6-7} \cmidrule(lr){8-9} \cmidrule(lr){10-11} \cmidrule(lr){12-13}
Method & LE & SSL\_1\% & LE & SSL\_1\% & LE & SSL\_1\% & LE & SSL\_1\% & LE & SSL\_1\% & LE & SSL\_1\%   \\ \hline
SimCLR \cite{pmlr_v119_chen20j}  & 43.80 & 40.81 & 35.00 & \underline{38.12} & 31.66 & \underline{27.60} & \underline{43.03} & 40.34 & 33.87 & \underline{35.68} & \underline{39.55} & \underline{47.83}    \\ 
SimSiam \cite{Chen_2021_CVPR} & 35.80 & \underline{47.55} & 14.78 & 34.80 & 19.87 & 22.87 & 29.18 & 40.20 & 21.31 & 32.60 & 26.46 & 45.24    \\
Dino \cite{Caron_2021_ICCV} & \underline{44.91} & 43.27 & \underline{36.78} & 33.02 & \underline{33.33} & 27.48 & 41.84 & \underline{42.69} & \underline{34.73} & 31.35 & 39.33 & 45.41  \\ 
Our & \textbf{52.88} & \textbf{48.15} & \textbf{49.65} & \textbf{46.62} & \textbf{42.94} & \textbf{29.98} & \textbf{54.18} & \textbf{48.14} & \textbf{39.32} & \textbf{36.94} & \textbf{44.02} & \textbf{48.62}  \\
\hline \hline
\multicolumn{9}{l}{\textbf{bold} denotes the best, \underline{underline} denotes the second best.}
\end{tabular}
\label{tab3}
\end{table*}

\begin{table}[h]
\small
\renewcommand\arraystretch{1.3}
\centering
\caption{The result of transfer learning with multiple source transfer setting. }
\begin{tabular}{lcccc}
\hline\hline
Source Dataset & \multicolumn{4}{c}{HKDD\_AMC36 \& HisarMod2019.1 (\(\%\))}  \\ \cmidrule(lr){2-5}
Target Dataset & \multicolumn{2}{c}{RML2016.10A  } & \multicolumn{2}{c}{HKDD\_AMC12} \\ \cmidrule(lr){2-3} \cmidrule(lr){4-5}
Method & LE & SSL\_1\% & LE & SSL\_1\%   \\ \hline
SimCLR \cite{pmlr_v119_chen20j} & 42.94 & 39.22 & 32.42 & \underline{37.88}       \\ 
SimSiam \cite{Chen_2021_CVPR} & 37.93 & \underline{46.48} & 19.44 & 34.42     \\
Dino \cite{Caron_2021_ICCV} & \underline{49.28} & 43.30 & \underline{38.88} & 35.60  \\ 
Our & \textbf{54.32} & \textbf{48.97} & \textbf{48.15} & \textbf{44.46}    \\
\hline \hline
\multicolumn{5}{l}{\textbf{bold} denotes the best, \underline{underline} denotes the second best.}
\end{tabular}
\label{tab4}
\end{table}

We evaluate the transfer learning performance across four datasets in both linear evaluation (LE) and semi-supervised learning with 1\% labeled data (SSL\_1\%)). Specifically, we first constructed a single-source transfer setting, where pre-trained the model on a single source dataset and then transferred to the remaining target dataset. On top of this setting, we additionally construct a multi-source transfer setting in which multiple datasets are jointly used as the source domain, enabling us to examine whether a richer and more diverse data distribution leads to improved generalization.

As shown in Table \ref{tab3}, under the single-source transfer setting, the proposed method achieves the best performance across all transfer directions. For example, when pretrained on HKDD\_AMC36 and transferred to RML2016.10A, proposed method obtains 52.88\% in linear evaluation, outperforming the best baseline by approximately 8\%, and maintains its superiority under the SSL\_1\% protocol as well. When the transfer targets dataset become more challenging, such as HKDD\_AMC12 and HisarMod2019.1, the performance gap becomes even more pronounced, with proposed method surpassing the strongest baseline by roughly 5\% to 10\% in both LE and SSL\_1\% protocol. When HisarMod2019.1 as the source dataset, the proposed method still outperforms all baseline methods on every target dataset under both LE and SSL\_1\% protocol. These results demonstrate that the representations learned by proposed method do not overfit a specific domain. Instead, they capture physically meaningful structural features, enabling strong discriminability even when transferred to simpler or distribution-shifted domains.

In the multi-source transfer setting, we jointly use HKDD\_AMC36 and HisarMod2019.1 for pretraining before transferring the model to RML2016.10A and HKDD\_AMC12. The results in Table \ref{tab4} show that multi-source transfer further enhances cross-domain generalization. On RML2016.10A, proposed method achieves 54.32\% in LE and 48.97\% in SSL\_1\%, exceeding the strongest baseline by approximately 5.1\% and 2.5\%, respectively. A similar performance gain is observed on HKDD\_AMC12. Notably, when compared with the single-source transfer results in Table \ref{tab3}, multi-source transfer provides consistent improvements across all transfer directions. For instance, on RML2016.10A, multi-source transfer improves LE and SSL\_1\% by 1.5\% and 0.8\%, respectively, over using HKDD\_AMC36 alone on single-source transfer setting. This indicates that by exposing the model to a larger-scale and more complex joint data distribution, the multi-source transfer setting enables the learning of more universal, domain-agnostic structural patterns. Moreover, because the proposed method explicitly constructs multi-view and complementary representations during pretraining, the inclusion of multiple source domains further strengthens this representation space, allowing the model to maintain stable separability and strong generalization when transferred to previously unseen domains.

\subsection{Ablation Study}

\begin{table}[t]
\small
\renewcommand\arraystretch{1.3}
\setlength{\tabcolsep}{4.5pt}
\centering
\caption{The average performance of the ablation study.}
\begin{tabular}{cccccccc}
\hline\hline
\multicolumn{4}{c}{} & \multicolumn{2}{c}{HKDD\_AMC12} & \multicolumn{2}{c}{HKDD\_AMC36} \\ \cmidrule(lr){5-6} \cmidrule(lr){7-8}
$\mathcal{L}_{ta}$ & $\mathcal{L}_{ap}$ & $\mathcal{L}_{fft}$ & $\mathcal{L}_{emd}$ & LE & SSL\_1\% & LE & SSL\_1\%  \\
\hline
\ding{51} & \ding{55} & \ding{55} & \ding{55} & 33.45 & 32.55 & 34.42 & 47.29 \\
\ding{55} & \ding{51} & \ding{55} & \ding{55} & 23.69 & 30.85 & 20.64 & 40.81 \\
\ding{55} & \ding{55} & \ding{51} & \ding{55} & 39.78 & 46.21 & 42.72 & 47.31 \\
\ding{55} & \ding{55} & \ding{55} & \ding{51} & 19.96 & 28.61 & 14.54 & 40.07 \\
\ding{55} & \ding{55} & \ding{51} & \ding{51} & \underline{47.48} & 46.24 & 47.94 & 49.29 \\
\ding{55} & \ding{51} & \ding{51} & \ding{51} & 45.01 & \textbf{48.26} & \textbf{48.62} & \underline{49.60} \\
\ding{51} & \ding{51} & \ding{51} & \ding{51} & \textbf{47.58} & \underline{48.16} & \underline{48.36} & \textbf{51.18} \\
\hline\hline
\multicolumn{8}{l}{\textbf{bold} denotes the best, \underline{underline} denotes the second best.}
\end{tabular}
\label{tab5}
\end{table}

To analyze the effects of each equivalent transformation branch, we conduct an ablation study on the time-domain branch \(\mathcal{L}_{ta}\), instantaneous-domain branch \(\mathcal{L}_{ap}\), frequency-domain branch \(\mathcal{L}_{fft}\), and time–frequency branch \(\mathcal{L}_{emd}\). On HKDD\_AMC12 and HKDD\_AMC36, we pretrain models under configurations that retain a single branch or different combinations of branches, and evaluate their performance under both LE and SSL\_1\% protocols. The results are summarized in Table \ref{tab5}.

The single-branch results show that none of the individual equivalent transformation is sufficient to support optimal performance. For example, on HKDD\_AMC12, when using only a single branch, the performance of LE is 33.45\%, 23.69\%, 39.78\%, and 19.96\%, respectively, and the performance of SSL\_1\% is 32.55\%, 30.85\%, 46.21\%, and 28.61\%, respectively. All of these are substantially lower than the performance of full-branch, which achieves 47.58\% in LE and 48.16\% in SSL\_1\%. Similarly, the degradation becomes more pronounced on HKDD\_AMC36. When only the time–frequency branch is retained, the performance of LE drops to merely 14.54\%, and even the best-performing single branch \(\mathcal{L}_{fft}\) achieves only 42.72\% in LE and 47.31\% in SSL\_1\%. Both remain more than 5\% lower than the performance of full-branch, which reaches 48.36\% in LE and 51.18\% in SSL\_1\%. This indicates that while frequency domain feature provides strong discriminative capability, relying on a single view is insufficient to capture the diversity of signal characteristics.

When multiple branches are enabled simultaneously, substantial performance gains are observed. Notably, combining the frequency-domain \(\mathcal{L}_{fft}\) and time–frequency branches \(\mathcal{L}_{emd}\) yields 47.48\% in LE and 46.24\% in SSL\_1\% on HKDD\_AMC12 dataset, approaching the performance of full-branch. On the HKDD\_AMC36 dataset, the performances are 47.94\% in LE and 49.29\% in SSL\_1\%, respectively, representing at least a 5\% improvement compared to using only the frequency domain branch \(\mathcal{L}_{fft}\). These results demonstrate that the four branches exhibit strong complementarity. When all equivalent transformation branches are jointly employed, the model achieves the best overall performance on both datasets, indicating that the multi-view equivalent transformations impose a more complete and robust structural constraint on the representation space.

\section{Conclusion}
\label{conclusion}
In this paper, we propose an unsupervised equivalent contrastive learning method for radio signal recognition. The proposed method exploits large-scale unlabeled radio signals by constructing information-lossless and complementary equivalent representations in the time, instantaneous, frequency, and time–frequency domains, which are treated as multi-view inputs for equivalent contrastive learning. By employing an equivalant contrastive learning strategy, the learned features capture latent signal structures and exhibit improved robustness and transferability across radio signal recognition tasks. Moreover, since the fine-tuning stage operates directly on the original IQ signals without requiring additional equivalent transformations, the computational overhead of downstream adaptation is reduced. In future work, we will investigate incorporating richer physical priors and alternative similarity metrics to further refine the equivalent transformation design and strengthen representation learning.


\small
\bibliographystyle{IEEEtran}
\bibliography{bare_jrnl}

\begin{thebibliography}{10}
\providecommand{\url}[1]{#1}
\csname url@samestyle\endcsname
\providecommand{\newblock}{\relax}
\providecommand{\bibinfo}[2]{#2}
\providecommand{\BIBentrySTDinterwordspacing}{\spaceskip=0pt\relax}
\providecommand{\BIBentryALTinterwordstretchfactor}{4}
\providecommand{\BIBentryALTinterwordspacing}{\spaceskip=\fontdimen2\font plus
\BIBentryALTinterwordstretchfactor\fontdimen3\font minus
  \fontdimen4\font\relax}
\providecommand{\BIBforeignlanguage}[2]{{%
\expandafter\ifx\csname l@#1\endcsname\relax
\typeout{** WARNING: IEEEtran.bst: No hyphenation pattern has been}%
\typeout{** loaded for the language `#1'. Using the pattern for}%
\typeout{** the default language instead.}%
\else
\language=\csname l@#1\endcsname
\fi
#2}}
\providecommand{\BIBdecl}{\relax}
\BIBdecl

\bibitem{9711564}
M.~Vaezi, A.~Azari, S.~R. Khosravirad, M.~Shirvanimoghaddam, M.~M. Azari,
  D.~Chasaki, and P.~Popovski, ``Cellular, wide-area, and non-terrestrial iot:
  A survey on 5g advances and the road toward 6g,'' \emph{IEEE Communications
  Surveys \& Tutorials}, vol.~24, no.~2, pp. 1117--1174, 2022.

\bibitem{8758230}
N.~Wang, P.~Wang, A.~Alipour-Fanid, L.~Jiao, and K.~Zeng, ``Physical-layer
  security of 5g wireless networks for iot: Challenges and opportunities,''
  \emph{IEEE Internet of Things Journal}, vol.~6, no.~5, pp. 8169--8181, 2019.

\bibitem{chaccour2022seven}
C.~Chaccour, M.~N. Soorki, W.~Saad, M.~Bennis, P.~Popovski, and M.~Debbah,
  ``Seven defining features of terahertz (thz) wireless systems: A fellowship
  of communication and sensing,'' \emph{IEEE Communications Surveys \&
  Tutorials}, vol.~24, no.~2, pp. 967--993, 2022.

\bibitem{11175176}
C.~Li, M.~Dong, Y.~Fu, F.~Richard~Yu, and N.~Cheng, ``Integrated sensing,
  communication, and computation for iov: Challenges and opportunities,''
  \emph{IEEE Communications Surveys \& Tutorials}, pp. 1--1, 2025.

\bibitem{10485272}
N.~Ye, S.~Miao, J.~Pan, Q.~Ouyang, X.~Li, and X.~Hou, ``Artificial intelligence
  for wireless physical-layer technologies (ai4phy): A comprehensive survey,''
  \emph{IEEE Transactions on Cognitive Communications and Networking}, vol.~10,
  no.~3, pp. 729--755, 2024.

\bibitem{liang2011cognitive}
Y.-C. Liang, K.-C. Chen, G.~Y. Li, and P.~Mahonen, ``Cognitive radio networking
  and communications: An overview,'' \emph{IEEE transactions on vehicular
  technology}, vol.~60, no.~7, pp. 3386--3407, 2011.

\bibitem{he2023channel}
J.~He, S.~Huang, Z.~Yang, K.~Yu, H.~Huan, and Z.~Feng, ``Channel-agnostic radio
  frequency fingerprint identification using spectral quotient constellation
  errors,'' \emph{IEEE Transactions on Wireless Communications}, vol.~23,
  no.~1, pp. 158--170, 2023.

\bibitem{dobre2007survey}
O.~A. Dobre, A.~Abdi, Y.~Bar-Ness, and W.~Su, ``Survey of automatic modulation
  classification techniques: classical approaches and new trends,'' \emph{IET
  communications}, vol.~1, no.~2, pp. 137--156, 2007.

\bibitem{4600222}
H.-C. Wu, M.~Saquib, and Z.~Yun, ``Novel automatic modulation classification
  using cumulant features for communications via multipath channels,''
  \emph{IEEE Transactions on Wireless Communications}, vol.~7, no.~8, pp.
  3098--3105, 2008.

\bibitem{5351708}
F.~Hameed, O.~A. Dobre, and D.~C. Popescu, ``On the likelihood-based approach
  to modulation classification,'' \emph{IEEE Transactions on Wireless
  Communications}, vol.~8, no.~12, pp. 5884--5892, 2009.

\bibitem{zhu2018likelihood}
D.~Zhu, V.~J. Mathews, and D.~H. Detienne, ``A likelihood-based algorithm for
  blind identification of qam and psk signals,'' \emph{IEEE Transactions on
  Wireless Communications}, vol.~17, no.~5, pp. 3417--3430, 2018.

\bibitem{TU202235}
Y.~Tu, Y.~Lin, H.~Zha, J.~Zhang, Y.~Wang, G.~Gui, and S.~Mao, ``Large-scale
  real-world radio signal recognition with deep learning,'' \emph{Chinese
  Journal of Aeronautics}, vol.~35, no.~9, pp. 35--48, 2022.

\bibitem{o2018over}
T.~J. O’Shea, T.~Roy, and T.~C. Clancy, ``Over-the-air deep learning based
  radio signal classification,'' \emph{IEEE Journal of Selected Topics in
  Signal Processing}, vol.~12, no.~1, pp. 168--179, 2018.

\bibitem{o2016convolutional}
T.~J. O’Shea, J.~Corgan, and T.~C. Clancy, ``Convolutional radio modulation
  recognition networks,'' proc. Int. Conf. Eng. Appl. Neural Netw. (EANN),
  213--226 (Springer, 2016).

\bibitem{9106397}
J.~Xu, C.~Luo, G.~Parr, and Y.~Luo, ``A spatiotemporal multi-channel learning
  framework for automatic modulation recognition,'' \emph{IEEE Wireless
  Communications Letters}, vol.~9, no.~10, pp. 1629--1632, 2020.

\bibitem{10146312}
K.~Qiu, S.~Zheng, L.~Zhang, C.~Lou, and X.~Yang, ``Deepsig: A hybrid
  heterogeneous deep learning framework for radio signal classification,''
  \emph{IEEE Transactions on Wireless Communications}, vol.~23, no.~1, pp.
  775--788, 2024.

\bibitem{10559458}
J.~Gui, T.~Chen, J.~Zhang, Q.~Cao, Z.~Sun, H.~Luo, and D.~Tao, ``A survey on
  self-supervised learning: Algorithms, applications, and future trends,''
  \emph{IEEE Transactions on Pattern Analysis and Machine Intelligence},
  vol.~46, no.~12, pp. 9052--9071, 2024.

\bibitem{hu2024comprehensive}
H.~Hu, X.~Wang, Y.~Zhang, Q.~Chen, and Q.~Guan, ``A comprehensive survey on
  contrastive learning,'' \emph{Neurocomputing}, vol. 610, p. 128645, 2024.

\bibitem{pmlr_v119_chen20j}
T.~Chen, S.~Kornblith, M.~Norouzi, and G.~Hinton, ``A simple framework for
  contrastive learning of visual representations,'' international Conference on
  Machine Learning (ICML), 1597--1607 (PMLR, 2020).

\bibitem{He_2020_CVPR}
K.~He, H.~Fan, Y.~Wu, S.~Xie, and R.~Girshick, ``Momentum contrast for
  unsupervised visual representation learning,'' iEEE/CVF Conference on
  Computer Vision and Pattern Recognition (CVPR), 9726-9735 (IEEE 2020).

\bibitem{Caron_2021_ICCV}
M.~Caron, H.~Touvron, I.~Misra, H.~J\'egou, J.~Mairal, P.~Bojanowski, and
  A.~Joulin, ``Emerging properties in self-supervised vision transformers,''
  iEEE/CVF International Conference on Computer Vision (ICCV), 9650-9660 (IEEE
  2021).

\bibitem{10093837}
W.~Kong, X.~Jiao, Y.~Xu, B.~Zhang, and Q.~Yang, ``A transformer-based
  contrastive semi-supervised learning framework for automatic modulation
  recognition,'' \emph{IEEE Transactions on Cognitive Communications and
  Networking}, vol.~9, no.~4, pp. 950--962, 2023.

\bibitem{10857965}
M.~Du, J.~Pan, and D.~Bi, ``A contrastive learner for automatic modulation
  classification,'' \emph{IEEE Transactions on Wireless Communications},
  vol.~24, no.~4, pp. 3575--3589, 2025.

\bibitem{10382665}
J.~Bai, X.~Wang, Z.~Xiao, H.~Zhou, T.~A.~A. Ali, Y.~Li, and L.~Jiao,
  ``Achieving efficient feature representation for modulation signal: A
  cooperative contrast learning approach,'' \emph{IEEE Internet of Things
  Journal}, vol.~11, no.~9, pp. 16\,196--16\,211, 2024.

\bibitem{li2025multi}
Y.~Li, X.~Shi, H.~Tan, Z.~Zhang, X.~Yang, and F.~Zhou, ``Multi-representation
  domain attentive contrastive learning based unsupervised automatic modulation
  recognition,'' \emph{Nature Communications}, vol.~16, no.~1, p. 5951, 2025.

\bibitem{9451544}
Z.~Li, F.~Liu, W.~Yang, S.~Peng, and J.~Zhou, ``A survey of convolutional
  neural networks: Analysis, applications, and prospects,'' \emph{IEEE
  Transactions on Neural Networks and Learning Systems}, vol.~33, no.~12, pp.
  6999--7019, 2022.

\bibitem{vaswani2017attention}
A.~Vaswani, N.~Shazeer, N.~Parmar, J.~Uszkoreit, L.~Jones, A.~N. Gomez,
  {\L}.~Kaiser, and I.~Polosukhin, ``Attention is all you need,''
  \emph{Advances in neural information processing systems}, vol.~30, 2017.

\bibitem{NEURIPS2020_70feb62b}
M.~Caron, I.~Misra, J.~Mairal, P.~Goyal, P.~Bojanowski, and A.~Joulin,
  ``Unsupervised learning of visual features by contrasting cluster
  assignments,'' advances in Neural Information Processing Systems (NeurIPS),
  9912--9924 (Springer 2020).

\bibitem{NEURIPS2020_f3ada80d}
J.-B. Grill, F.~Strub, F.~Altch\'{e}, C.~Tallec, P.~Richemond, E.~Buchatskaya,
  C.~Doersch, B.~Avila~Pires, Z.~Guo, M.~Gheshlaghi~Azar, B.~Piot,
  k.~kavukcuoglu, R.~Munos, and M.~Valko, ``Bootstrap your own latent - a new
  approach to self-supervised learning,'' advances in Neural Information
  Processing Systems (NeurIPS), 21271--21284 (Springer 2020).

\bibitem{Chen_2021_CVPR}
X.~Chen and K.~He, ``Exploring simple siamese representation learning,''
  iEEE/CVF Conference on Computer Vision and Pattern Recognition (CVPR),
  15750-15758 (IEEE 2021).

\bibitem{pmlr_v139_zbontar21a}
J.~Zbontar, L.~Jing, I.~Misra, Y.~LeCun, and S.~Deny, ``Barlow twins:
  Self-supervised learning via redundancy reduction,'' international Conference
  on Machine Learning (ICML), 12310--12320 (PMLR, 2021).

\bibitem{He_2022_CVPR}
K.~He, X.~Chen, S.~Xie, Y.~Li, P.~Doll\'ar, and R.~Girshick, ``Masked
  autoencoders are scalable vision learners,'' iEEE/CVF Conference on Computer
  Vision and Pattern Recognition (CVPR), 16000-16009 (IEEE 2022).

\bibitem{10562208}
C.~Xiao, S.~Yang, Z.~Feng, and L.~Jiao, ``Mclhn: Toward automatic modulation
  classification via masked contrastive learning with hard negatives,''
  \emph{IEEE Transactions on Wireless Communications}, vol.~23, no.~10, pp.
  14\,304--14\,319, 2024.

\bibitem{heckbert1995fourier}
P.~Heckbert, ``Fourier transforms and the fast fourier transform (fft)
  algorithm,'' \emph{Computer Graphics}, vol.~2, no. 1995, pp. 15--463, 1995.

\bibitem{huang1998empirical}
N.~E. Huang, Z.~Shen, S.~R. Long, M.~C. Wu, H.~H. Shih, Q.~Zheng, N.-C. Yen,
  C.~C. Tung, and H.~H. Liu, ``The empirical mode decomposition and the hilbert
  spectrum for nonlinear and non-stationary time series analysis,''
  \emph{Proceedings of the Royal Society of London. Series A: mathematical,
  physical and engineering sciences}, vol. 454, no. 1971, pp. 903--995, 1998.

\bibitem{10042021}
S.~Zheng, X.~Zhou, L.~Zhang, P.~Qi, K.~Qiu, J.~Zhu, and X.~Yang, ``Toward
  next-generation signal intelligence: A hybrid knowledge and data-driven deep
  learning framework for radio signal classification,'' \emph{IEEE Transactions
  on Cognitive Communications and Networking}, vol.~9, no.~3, pp. 564--579,
  2023.

\bibitem{zhang2022deep}
F.~Zhang, C.~Luo, J.~Xu, Y.~Luo, and F.-C. Zheng, ``Deep learning based
  automatic modulation recognition: Models, datasets, and challenges,''
  \emph{Digital Signal Processing}, vol. 129, p. 103650, 2022.

\bibitem{xie2017aggregated}
S.~Xie, R.~Girshick, P.~Doll{\'a}r, Z.~Tu, and K.~He, ``Aggregated residual
  transformations for deep neural networks,'' iEEE/CVF Conference on Computer
  Vision and Pattern Recognition (CVPR), 1492--1500 (IEEE 2017).

\bibitem{zhai2019s4l}
X.~Zhai, A.~Oliver, A.~Kolesnikov, and L.~Beyer, ``S4l: Self-supervised
  semi-supervised learning,'' iEEE/CVF Conference on Computer Vision and
  Pattern Recognition (CVPR), 1476--1485 (IEEE 2019).

\end{thebibliography}

\end{document}